\documentclass{PoS}

\title{The nature of the Roberge-Weiss Transition in $N_f=2$ QCD with Wilson Fermions on $N_\tau=6$ lattices}

\ShortTitle{The nature of the Roberge-Weiss Transition in $N_f=2$ QCD with Wilson Fermions on $N_\tau=6$ lattices}

\author{Francesca Cuteri$^1$, \speaker{Christopher Czaban}$^{1,2}$, Owe Philipsen$^{1,2}$, Christopher Pinke$^1$, Alessandro Sciarra$^1$\\
        \llap{$^1$}Institut für Theoretische Physik - Johann Wolfgang Goethe-Universität\\
        Max-von-Laue-Str. 1, 60438 Frankfurt am Main\\        
		\llap{$^2$}John von Neumann Institute for Computing (NIC)\\
		GSI, Planckstr. 1, 64291 Darmstadt, Germany \\
		E-mail: \email{cuteri, czaban, philipsen, pinke, sciarra @th.physik.uni-frankfurt.de} \\
		}

\abstract{
The finite temperature chiral and deconfinement phase transitions 
at zero density for light and heavy quarks, respectively, have
analytic continuations to imaginary chemical potential. At some 
critical imaginary chemical potential, they meet the 
Roberge-Weiss transition between adjacent $Z(3)$ sectors.
For light and heavy 
quarks, where the chiral and deconfinement transitions are first order, 
the transition lines meet in a triple point. 
For intermediate masses chiral or deconfinement transitions are crossover
and the Roberge-Weiss transition ends in a second order point.
At the boundary between these regimes the junction is
a tricritical point, 
as shown in 
studies with $N_f=2,3$ flavors of staggered and Wilson quarks on $N_\tau=4$ lattices. 
Employing finite size scaling we 
investigate the nature of this point as a function of quark 
mass for $N_f=2$ flavors of Wilson fermions with a temporal lattice extent 
of $N_\tau=6$. In particular we are interested in the change of the 
location of tricritical points compared to our ealier
study on $N_\tau=4$.
}

\FullConference{The 33rd International Symposium on Lattice Field Theory\\
                 14 -18 July  2015\\
                 Kobe International Conference Center, Kobe, Japan}

\usepackage[utf8]{inputenc}
\usepackage[T1]{fontenc}
\usepackage{amsmath, amssymb}
\usepackage[format=hang, font=small,labelfont=bf,textfont=it, justification=centerlast, margin=1mm]{caption}
\usepackage{dsfont}
\usepackage{booktabs}

\newcommand{\Tr}{\operatorname{Tr}}
\def\expect#1{\ensuremath{\left\langle{#1}\right\rangle}}
\newcommand{\abs}[1]{\lvert {#1}\,\rvert} 
\newcommand{\ee}{\ensuremath{\textrm{e}}}

\newcommand{\Loewe}{LOEWE-CSC}
\newcommand{\Lcsc}{L-CSC}
\newcommand{\clqcd}{CL\kern-.25em\textsuperscript{2}QCD}

\begin{document}

\section{Introduction}
\begin{figure}[t]
\begin{minipage}{0.49\textwidth}
\center
\vspace{-0.3cm}
\includegraphics[width=0.8\linewidth]{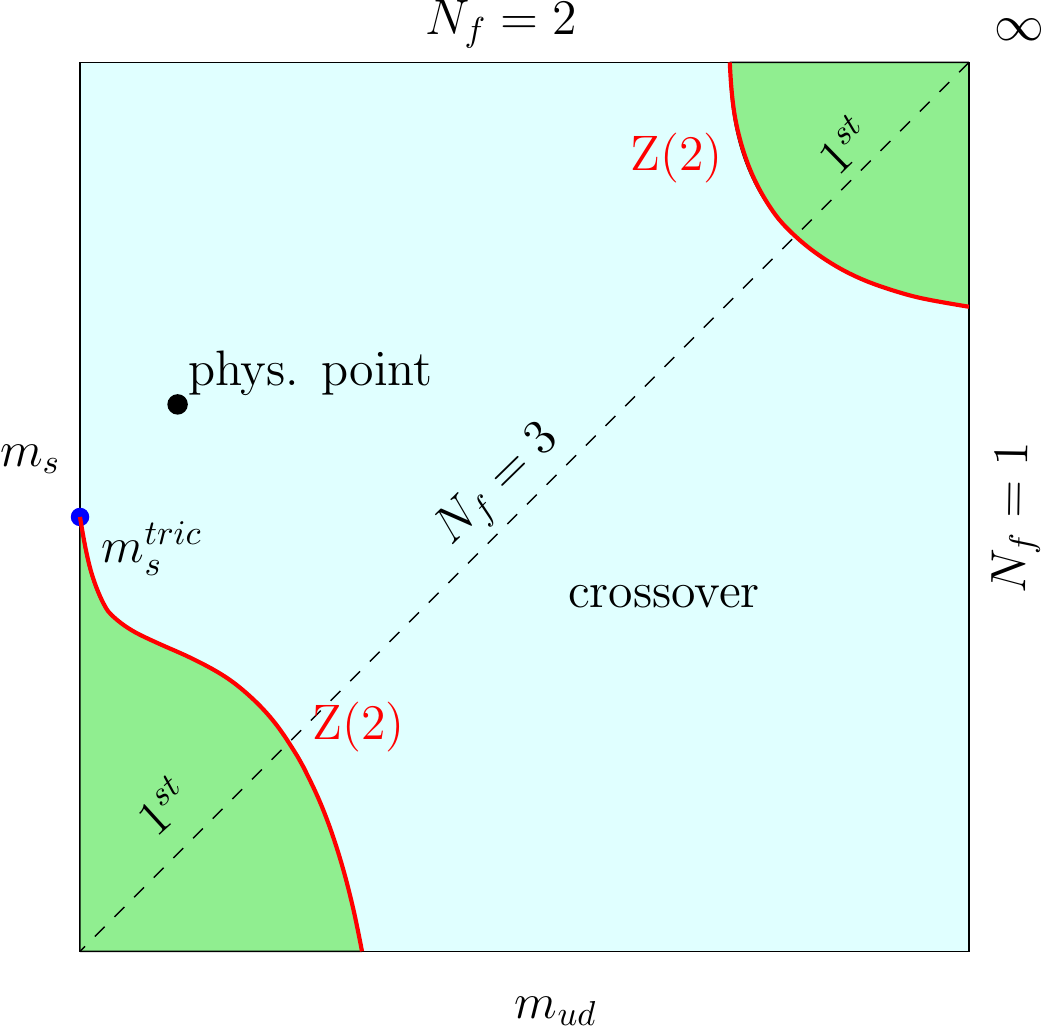}
\end{minipage}
\begin{minipage}{0.49\textwidth}
\center
\includegraphics[width=0.75\linewidth]{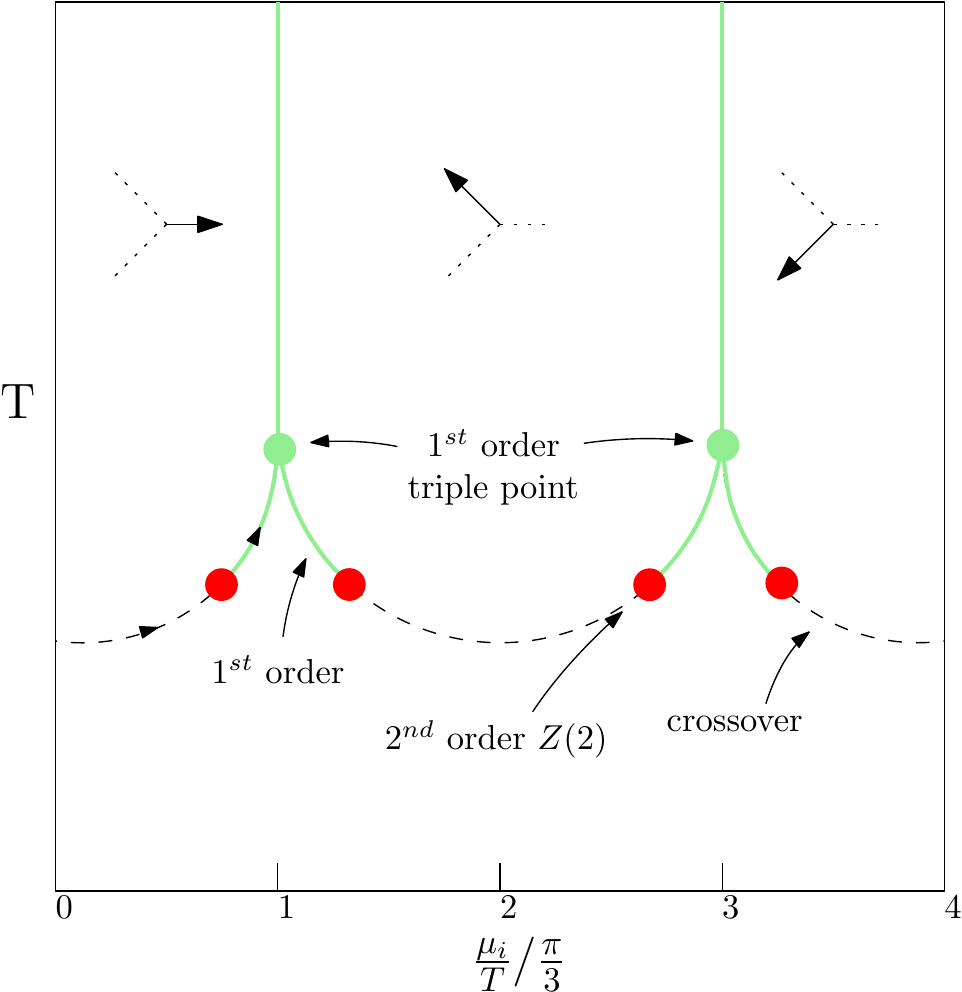}
\end{minipage}
\caption{(Left) Columbia plot at $\mu=0$. (Right) Schematic phase diagram of QCD at imaginary chemical potential. Illustrated is the change of the nature of the chiral/deconf. transition for a particular combination of $m_{u,d}$, $m_s$.}
\label{Fig:ColumbiaPlot}
\end{figure}
The QCD phase diagram confronts us with many open questions. 
An interesting one addresses the nature of the phase transition in the two flavor chiral limit 
at zero chemical potential which has yet to be determined (see Fig. \ref{Fig:ColumbiaPlot}) and goes along with important insights 
about the physical QCD phase diagram. 
Symmetry arguments rule out an analytic crossover leaving either a 
first order or a second order phase transition. In the past many attempts have been undertaken 
to map out this region with standard simulation techniques resulting in many contradicting \cite{Iwasaki:1996ya,D'Elia:2005bv,Ejiri:2009ac}
lattice results between different fermion discretizations (e.g. staggered and Wilson). 
Closely related to this issue are the high computational costs arising due to low quark masses. 
A different approach to simulating at increasingly smaller quark masses  
is to choose a purely imaginary chemical potential (proposed in \cite{Bonati:2014kpa}) for which hybrid Monte Carlo (HMC) 
algorithms can be applied due to the absence of the sign problem. 
For purely imaginary chemical potential the center symmetry is extended to the periodic 
Roberge-Weiss symmetry \cite{Roberge:1986mm}. In this region QCD exhibits a rich phase structure 
and features properties that can be used to impose constraints on the phase diagram at real chemical 
potential to a certain extent. In former studies on $N_\tau=4$ lattices \cite{Philipsen:2014rpa} 
we investigated these 
properties for $N_f=2$ flavors of Wilson fermions by mapping out tricritical points at a 
critical value of $\mu_i$. The present work presents our results 
of progressing to a finer lattice with an temporal extent of $N_\tau=6$, taking a step towards the 
continuum.  
\section{The QCD phase structure for imaginary $\mu$}
\begin{figure}[t]
\begin{minipage}{0.49\textwidth}
\center
\includegraphics[width=1.1\linewidth]{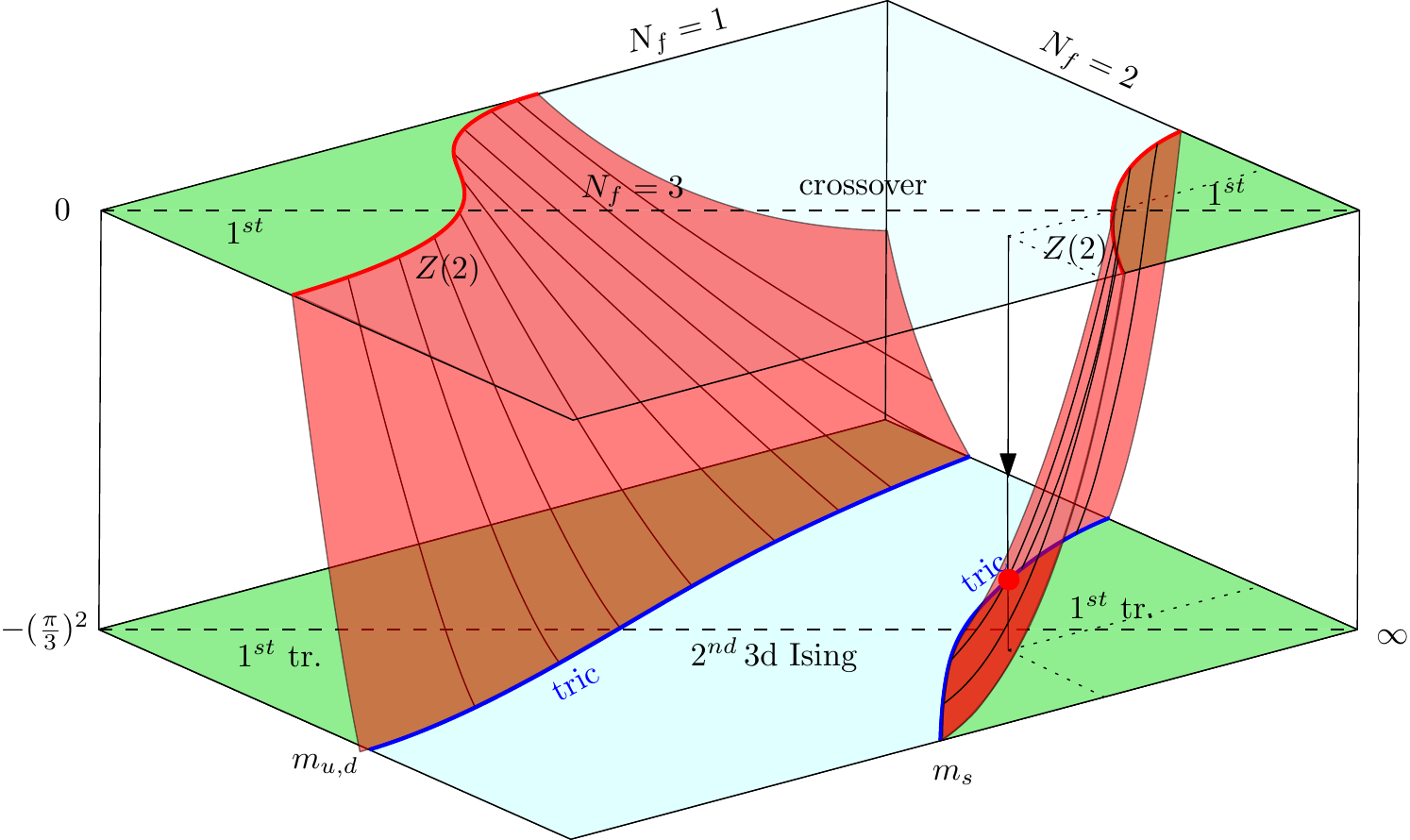}
\end{minipage}
\begin{minipage}{0.49\textwidth}
\center
\includegraphics[width=0.8\linewidth]{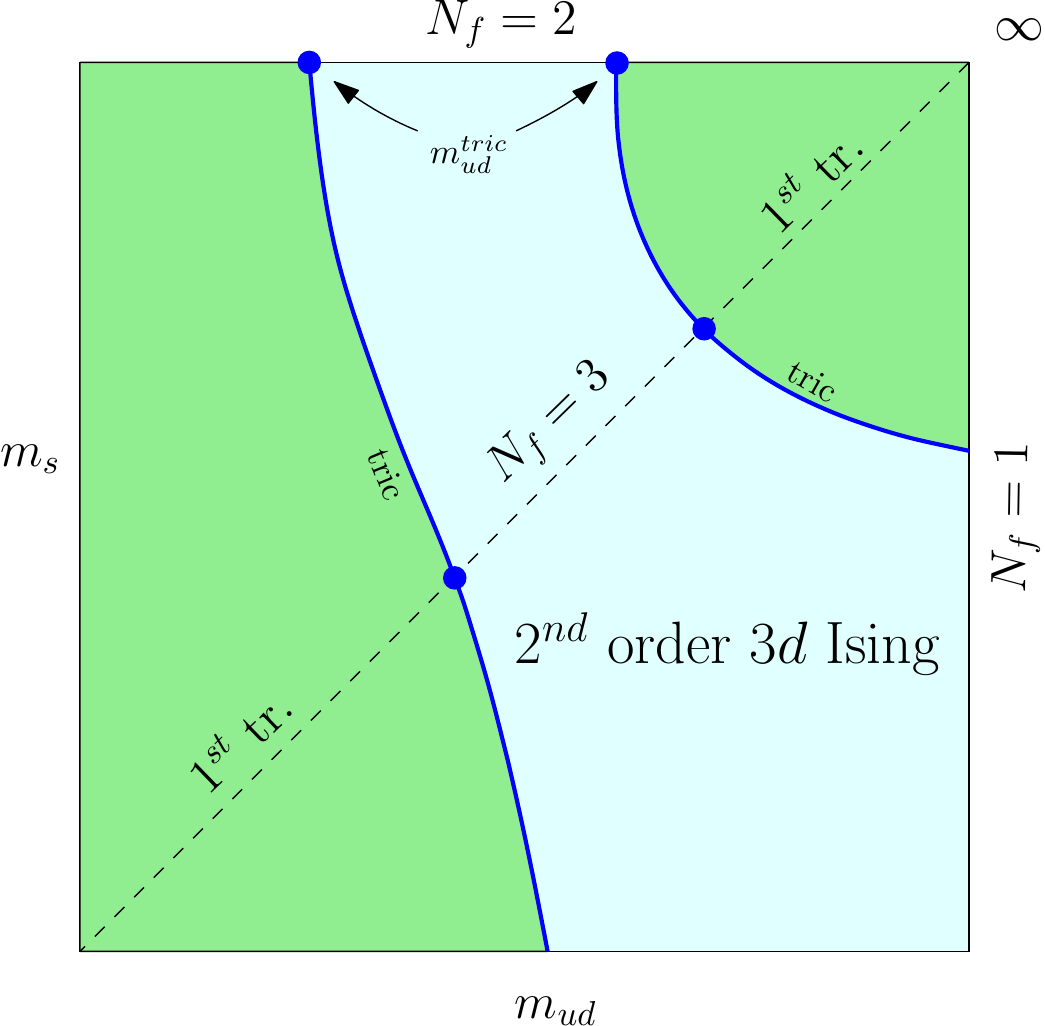}
\end{minipage}
\caption{(Left) The Columbia plot with $(\mu/T)^2$ as an additional parameter. The $Z(2)$ critical lines became surfaces terminating
in tricritical lines at critical values $(\mu_i^c/T)$. (Right) A Stripe of the left figure at $(\mu_i^c/T)$.}
\label{Fig:ImagMuColumbiaPlot}
\end{figure}
For purely imaginary chemical potential the symmetries of the QCD partition function are reflection in $\mu$ 
and periodicity \cite{Roberge:1986mm}
\begin{align}
	Z(\mu) = Z(-\mu)\;, \qquad
	Z\left(\frac{\mu}{T} \right) =  Z\left( \frac{\mu}{T}+i\frac{2\pi n}{3} \right).
	\label{Eq:ExactSymmetries}
\end{align}
As shown in Fig. \ref{Fig:ColumbiaPlot} there are center sectors seperated by discrete values 
\begin{align}
	\left(\frac{\mu_i}{T}\right)_c = \frac{\pi}{3}(2n+1)\;,\; n \in \mathds{Z}.
	\label{Eq:RobergeWeissValue}
\end{align} 
Above a critical temperature $T_c$ the Roberge-Weiss (RW) transition between the $Z(3)$ center sectors is of first order, 
represented by the vertical lines, whereas it is a mere crossover below $T_c$ \cite{Roberge:1986mm,deForcrand:2002ci,D'Elia:2002gd}.
Physical observables are not sensitive to the sectors though it is possible to distinguish between the sectors via 
the phase $\varphi$ of the Polyakov loop $L$
\begin{align}
L({\bf n}) = \frac{1}{3} \Tr_C \left[ \prod_{n_0=0}^{N_\tau-1} U_0(n_0,{\bf n})\right] \equiv \abs{L({\bf n})}\ee^{i\varphi}.
\label{Eq:Polyakovloop}
\end{align}
Traversing the center sectors $\varphi$ takes on the values $\expect{\varphi} = 2n\pi/3$, $n \in \mathds{N}$.
Hence measurements of the Polyakov loop and plotting its real versus its imaginary part yields scatter plots that 
resemble the three-armed arrow symbols in Fig. \ref{Fig:ColumbiaPlot} (right).
As indicated by the dotted lines the chiral/deconfinement transition analytically 
continues from $\mu = 0$ to imaginary chemical potential until it joins the RW transition in its endpoint. At this junction $\mu_i$ takes on the critical values $(\mu_i/T)_c$ and the 
phase structure in the $m_{u,d}-m_s$ plane, shown 
in Fig. \ref{Fig:ImagMuColumbiaPlot} (right), is qualitatively different compared to zero chemical 
potential (c.f. Fig. \ref{Fig:ColumbiaPlot}).    
The nature of the chiral/deconfinement transition for $0 < \mu_i < (\mu_i/T)_c$ and of the RW endpoint is determined 
by the number of flavours and the values of the quark masses. Past studies \cite{Philipsen:2014rpa,Bonati:2014kpa} for $N_f=2,3$ have found the RW endpoints to be first order triple points for heavy and low masses and second order endpoints for intermediate masses.
This can be intuitively understood from drawing the columbia plot with the chemical potential as an additional 
parameter (see Fig. \ref{Fig:ImagMuColumbiaPlot}). The $Z(2)$ critical lines of $\mu=0$ become surfaces terminating in tricritical lines at
imaginary chemical potential $(\mu_i/T)_c$ (the RW endpoints).
Fixing in Fig. \ref{Fig:ImagMuColumbiaPlot} the value of the quark masses and varying the chemical potential from
$(\mu/T)^2 = 0$ to $-(\pi/3)^2$, as indicated by the arrow-headed vertical line, we move along the chiral/deconfinement transition line
in the $T-\mu$ plane (Fig. \ref{Fig:ColumbiaPlot} (right)). This line changes its nature from crossover to first order with a second order point in between, 
joining the RW transition in its endpoint at $(\mu_i/T)_c$ rendering it a triple point.
Depending on the mass the nature of the RW endpoint can be tricritical or second order $Z(2)$. Accordingly the chiral/deconfinement transition line can be first order or purely crossover.  
\section{Simulation details}
The computational and algorithmic setup used for this work is similar to the former study about the 
RW transition on $N_\tau=4$ lattices \cite{Philipsen:2014rpa}, i.e. with the standard Wilson gauge action
and $N_f=2$ flavors of unimproved Wilson fermions. The bare fermion mass $m_{u,d}\equiv m$ is indirectly 
controlled via the 
hopping parameter $\kappa = (2(am+4))^{-1}$. The temperature $T=1/a(\beta)N_\tau$ on the lattice is adjusted via change of 
the lattice coupling parameter $\beta$.
For a fixed temporal lattice extent of $N_\tau=6$ and imaginary chemical potential of $\mu_i/T=\pi$
we scanned for the tricritical points $\kappa_\text{tric}^\text{light}$ and $\kappa_\text{tric}^\text{heavy}$ 
in $\kappa \in (0.1,\ldots,0.165)$. For each value simulations at three to four different spatial
lattice extents were done, where we kept $N_\sigma\geq16$ except for $\kappa=0.1625$ with $N_\sigma\geq12$, which implies an aspect ratio of approximately three or larger in the most cases.  
For every spatial lattice extent the temperature scans included $\sim15$ $\beta$ values. The number of 
HMC trajectories (unit length) amounts to $\sim40k-500k$ per $\beta$ value after at least $5k$ thermalization steps.
In order to accumulate statistics faster the runs were distributed on four chains per $\beta$. This also
helped to decide whether the statistics were large enough.
Throughout all simulations the acceptance rate was tuned to $\sim75\%$. For very small masses, i.e. $\kappa\geq 0.16$ 
the Hasenbusch trick \cite{Hasenbusch:2001ne} was applied in order to improve integrator stability.
Our observables are the Polyakov loop
and the chiral condensate which we measure on every trajectory.
The simulations were performed with the OpenCL based code {\clqcd } \cite{Bach:2012iw} which is highly optimized
for running on graphic processing units (GPUs) on \Loewe \cite{LoeweRef} at 
Goethe University Frankfurt and on \Lcsc \cite{LcscRef} at GSI Darmstadt.     
\section{Analysis}
\begin{figure}[t]
\begin{minipage}{0.49\textwidth}
\center
\includegraphics[trim=0cm 0cm 0cm 2.3cm, width=\linewidth, clip=true, width=1.0\linewidth]{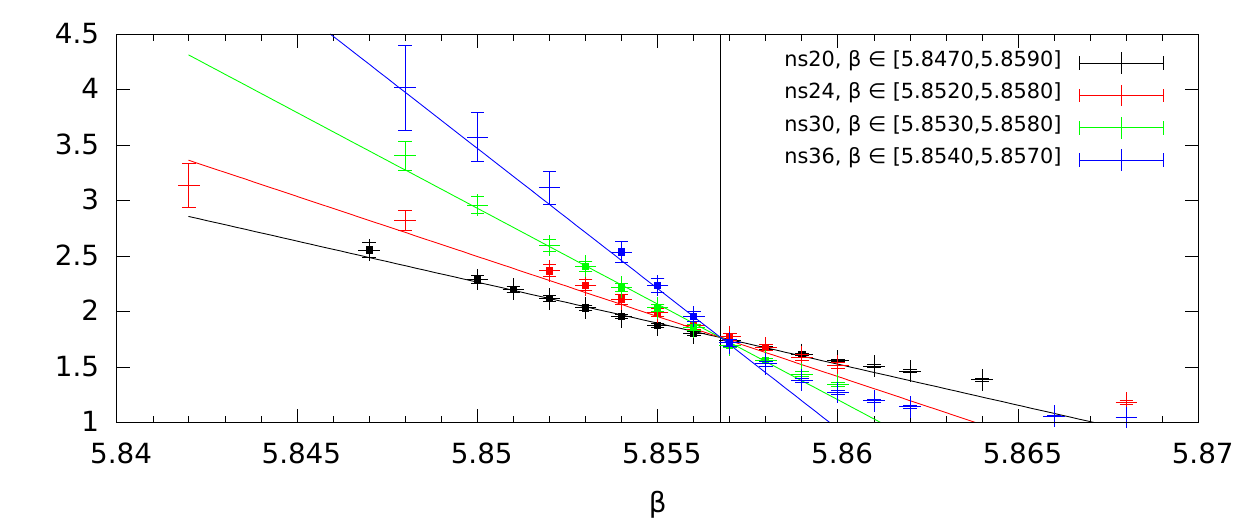}
\end{minipage}
\hspace{1cm}
\begin{minipage}{0.49\textwidth}
\center
\includegraphics[width=1.0\linewidth]{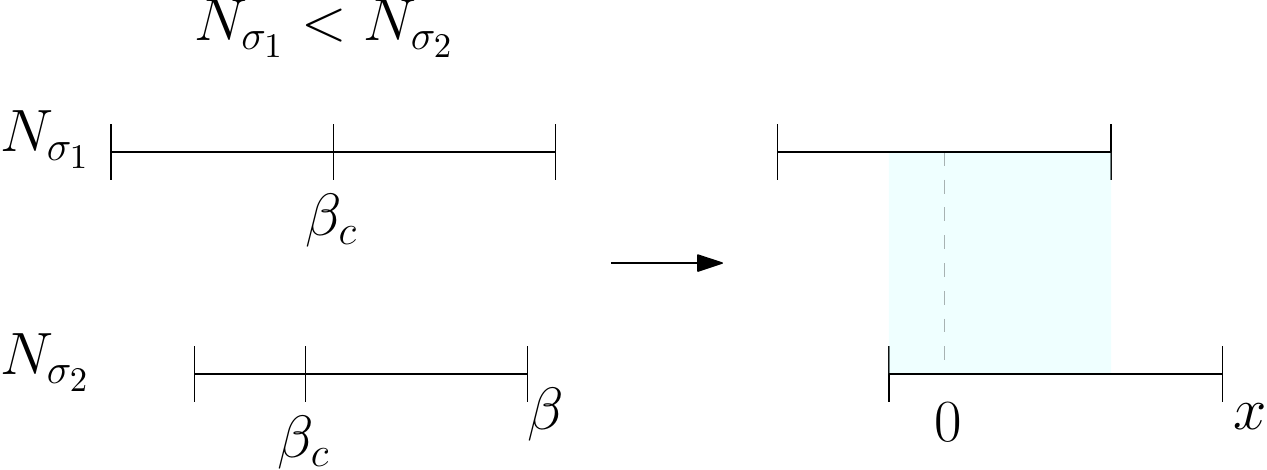}
\end{minipage}
\caption{(Left) $B_4$ as function of imaginary chemcial potential and $N_\sigma$ at a fixed quark mass values. The different curves intersect at the critical value. (Right) Translation of the fitting ranges from $\beta$ for different $N_\sigma$ to the scaling variable $x=(\beta-\beta_c)N_\sigma^{1/\nu}$.}
\label{Fig:FiniteSizeScaling}
\end{figure}

We estimate the autocorrelation on the observables by a python implementation of the Wolff method 
\cite{Wolff:2003sm}. An appropriate binning is then chosen in order to remove the autocorrelation effects in functions 
of the observables. In order to locate phase transitions and to extract their order we use the Binder cumulant
\begin{align}
    B_4(X,\alpha_1,\ldots,\alpha_n) = \frac{\expect{(X-\expect{X})^4}}{\expect{(X-\expect{X})^2}{}^2}, 
    \label{Eq:BinderCumulantDefinition}
\end{align}
whose values which are listed in table \ref{tab:BinderValues} indicate the order of the phase transition.
For $V\rightarrow\infty$ it is a non-analytic step function 
that depends on a set of 
parameters which are $\{\alpha_i\} = \{\beta,\mu_i^c\}$. We have chosen $X = L_{Im}$, 
where $L$ is the spatial average of the Polyakov loop (c.f. Eq.(\ref{Eq:Polyakovloop})). 
On finite volumes $B_4$ gets smoothed out, passing continuously through the critical value, with its slope increasing 
with the spatial lattice extent. Close to the critical value of $\beta$, $B_4$ is a function of $x\equiv(\beta-\beta_c)N_\sigma^{1/\nu}$ solely and due to its scaling behaviour in the vicinity of a critical point can be expanded around $0$ to leading order
\begin{align}
    B_4(\beta,N_\sigma) = B_4(\beta_c,\infty) + a_1x + a_2x^2 + \ldots
    \label{Eq:BinderCumulantExpansion}
\end{align}
\begin{table}[b]
  \centering
  \[
  \begin{array}{*{5}{c}}
	\toprule[0.3mm]
	& \text{Crossover} & 1^{st} \text{ triple} & \text{Tricritical} & 3D \text{ Ising} \\
	\midrule[0.1mm]
	B_4      & 3 & 1.5 & 2   & 1.604     \\
	\nu          & - & 1/3 & 1/2 & 0.6301(4) \\
	\gamma       & - & 1   & 1   & 1.2372(5) \\
	\bottomrule[0.3mm]
  \end{array}
  \]
  \caption{Critical values of $\nu$, $\gamma$ and $B_4\equiv B_4(X,\alpha_c)$
  		   for some universality classes 
  		  }
  \label{tab:BinderValues}
\end{table}     
Varying the lattice coupling $\beta$ we 
then measure $B_4$ at fixed $\mu_i$ along the phase boundary separating the RW sectors in the $T$-direction 
where the third moment of the fluctuations vanishes, $\expect{(X-\expect{X})^3} = 0$.  
This procedure is repeated for different spatial volumes each yielding a curve with a different slope. In order to smoothen
the raw measurements and to fill in additional points Ferrenberg-Swendsen reweighting 
\cite{FerrenbergSwendsenRef} was used. Subsequently a finite size
scaling study is employed by fitting Eq.(\ref{Eq:BinderCumulantExpansion}) to all data at once, as it is shown for an example 
in Fig. \ref{Fig:FiniteSizeScaling}. 
The fitted value $B_4(\beta_c,\infty)$ is observed to be always larger than the values listed in 
Eq.(\ref{Eq:BinderCumulantDefinition})(right) which is in agreement with previous observations 
\cite{deForcrand:2010he,Philipsen:2014rpa} and can be explained by finite
size effects not included in Eq.(\ref{Eq:BinderCumulantExpansion}). A thorough investigation of this issue will be included in our future article \cite{Sciarra}. 
Hence we concentrate on the critical exponent $\nu$ which is a better suited quantity since it is less prone to finite size effects. 
Close to a critical point it takes on its universal value depending on the universality class. 
The possible values are listed in table \ref{tab:BinderValues}.
Another important quantity is the susceptibility of the order parameter
\begin{align}
\chi(X) \equiv N_\sigma^3 \expect{(X-\expect{X})^2}, 
\end{align} 
which as well scales around the critical value of $\beta$ following
\begin{align}
\chi = N_\sigma^{\gamma/\nu} f(t N_\sigma^{1/\nu}),
\label{Eq:Collapse}
\end{align} 
with a universal scaling function $f$ and the reduced temperature $t\equiv(T-T_c)/T_c$.
This can be used to check for consistency by plotting $\chi/N_\sigma^{\gamma/\nu}$ against $tN_\sigma^{1/\nu}$.
All the different curves $\chi/N_\sigma^{\gamma/\nu}$ are supposed to collapse for a correctly determined value of $\nu$.  
A further remark addresses the fitting procedure of Eq.(\ref{Eq:BinderCumulantExpansion}) to the reweighted 
data $B_4(\beta_i,N_\sigma)$. 
Natural and straightforward criteria are $\chi^2\approx1$ and a goodness of the fit of $Q\approx50$. 
Relying solely on these does not yet guarantee meaningful results as long as there is not a reasonable overlap in 
$x=(\beta-\beta_c)N_\sigma^{1/\nu}$  and an approximate symmetry in the fitting range around $x=0$. 
The mapping from $\beta$ to $x$ is illustrated in Fig. \ref{Fig:FiniteSizeScaling}.
In order to respect these criteria we employ an algorithm which is repeating the fit for many combinations of $\beta$-ranges
of the different curves (which again is coupled to certain boundary conditions) and filters amongst all the fits with proper
$\chi^2$ and $Q$ the ones with the largest overlap in $x$ and symmetry around $x=0$. Again a detailed discussion 
about this will be included in our future article \cite{Sciarra}.
\section{Numerical results}
\begin{figure}[t]
\begin{minipage}{0.49\textwidth}
\center
\includegraphics[width=1.\linewidth]{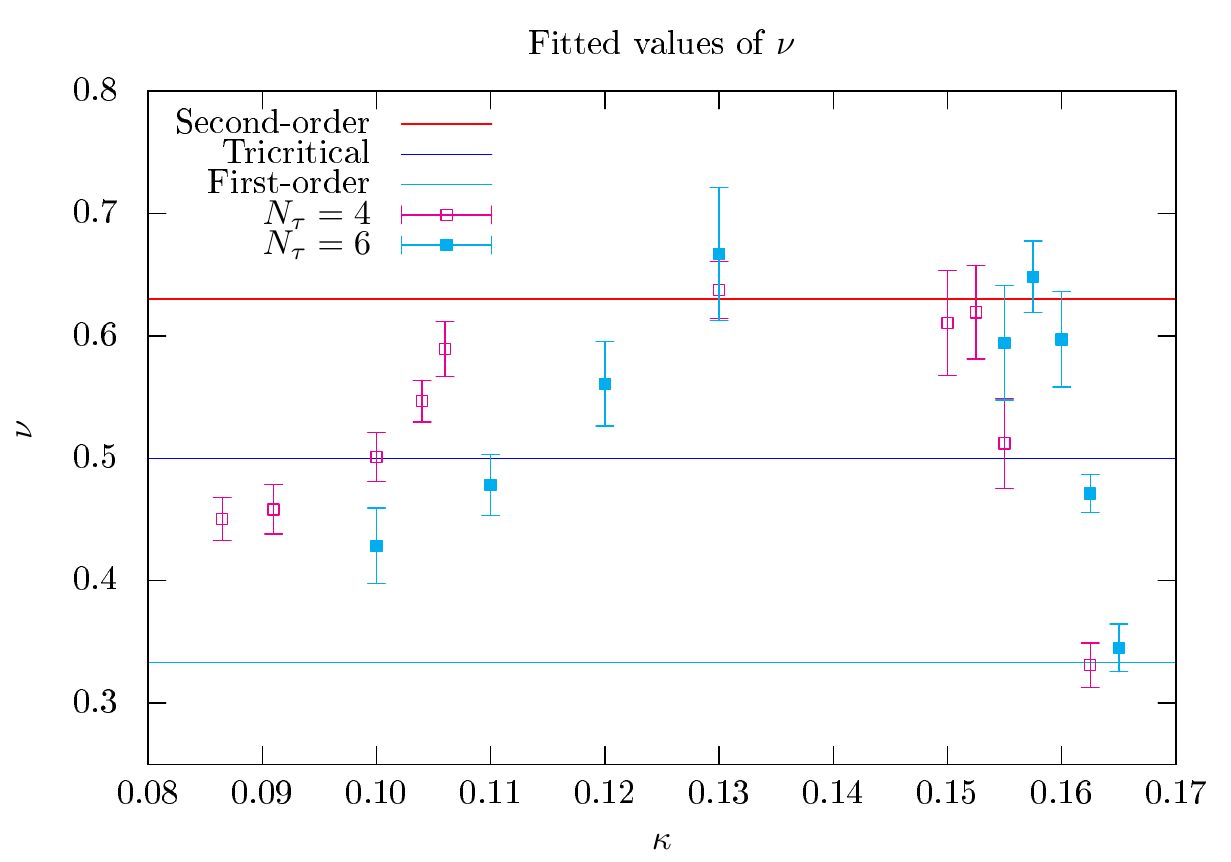}
\end{minipage}
\begin{minipage}{0.49\textwidth}
\center
\includegraphics[width=0.8\linewidth]{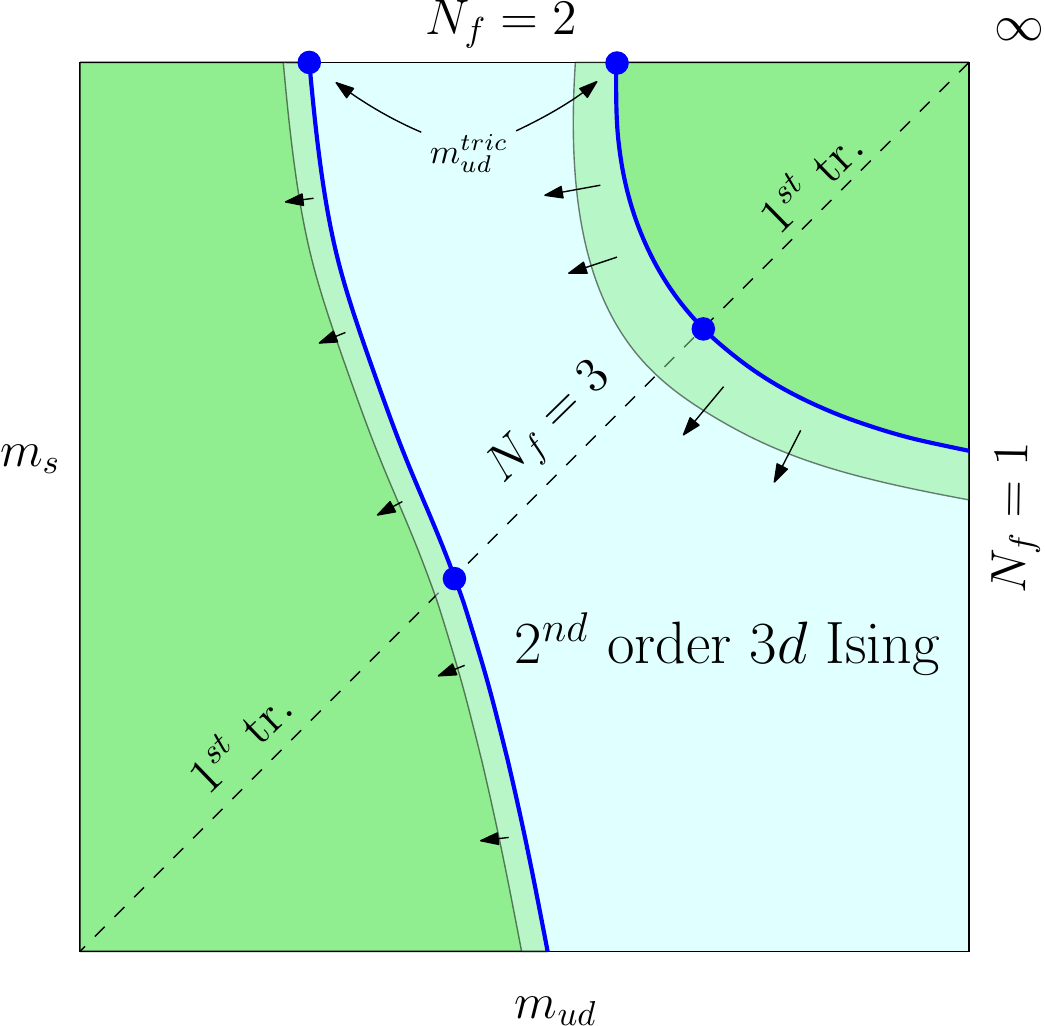}
\end{minipage}
\caption{(Left) Results of the $N_\tau=4,6$ study combined in one plot. The $N_\tau=6$ curve is shifted towards larger $\kappa$ values. (Right) Schematic of the shift of the tricritical lines at $\mu_i^c/T$ to larger $\kappa$ values.}
\label{Fig:Results}
\end{figure}

In the infinite volume limit the critical exponent is a step function of $\kappa$. This function
is smeared out on finite volumes as indicated in Fig. \ref{Fig:Results} (left). 
For this reason the collapse plot technique does not provide meaningful results for the case
when the critical exponent on the finite lattice is in between the values of table \ref{tab:BinderValues}.
Ultimately a further refinement is taken with the sophisticated fitting procedure briefly discussed in the previous section which 
allows for a more accurate resolution of the critical exponent $\nu$ in a rather quantitative way.
We estimate the following two critical values of $\kappa$
\begin{align}
\kappa^\text{tric}_\text{heavy} = 0.110(10) \text{\qquad and\qquad} \kappa^\text{tric}_\text{light} = 0.1625(25).
\end{align}  
In Fig. \ref{Fig:Results} (left) we compare the results of this work to the former study \cite{Philipsen:2014rpa} but
for the sake of clarity we did not include all the data points of the $N_\tau=4$ study. It appears
that the $N_\tau=6$ curve is shifted towards smaller mass values or a shift of the first order region 
towards lower mass values respectively as illustrated schematically in Fig. \ref{Fig:Results} (right).

\section{Summary and Perspectives}

QCD at imaginary chemical potential $\mu_i$ features many interesting properties which can be used to constrain the QCD phase
diagram at real $\mu$ where it is not possible to apply standard simulation algorithms due to the sign problem.
Our study of the imaginary region at $\mu_i=\pi T$ on $N_\tau=6$ lattices with unimproved Wilson 
fermions are consistent with
the findings of former studies and we observed the expected shift of the tricritical masses to smaller values. Despite getting one step closer to the continuum
it is still not possible to make a definite statement about the rate of the shift of $\kappa_\text{tric}$ in 
dependence of the temporal lattice extent. It will be interesting to see the results of the future studies on higher 
$N_\tau$ lattices.

\acknowledgments

This work is supported by the Helmholtz International Center for FAIR within the LOEWE program of the State of Hesse.
We thank the staff of \Loewe\ and \Lcsc\ at GU-Frankfurt and the NIC in J\"ulich for computer time and support.

\end{document}